\documentclass[10pt, conference, letterpaper]{IEEEtran}
\IEEEoverridecommandlockouts
\usepackage{cite}
\usepackage{amsmath,amssymb,amsthm}
\usepackage{bbm}
\usepackage{algpseudocode}
\usepackage{graphicx}
\usepackage{textcomp}
\usepackage{placeins}
\usepackage[russian,USenglish]{babel}
\usepackage{mathtools}
\usepackage{xcolor}
\usepackage{algorithm}
\usepackage{bm}
\usepackage{booktabs,threeparttable}
\usepackage{multirow}
\algnewcommand\algorithmicswitch{\textbf{switch}}
\algnewcommand\algorithmiccase{\textbf{case}}
\algdef{SE}[SWITCH]{Switch}{EndSwitch}[1]{\algorithmicswitch\ #1\ \algorithmicdo}{\algorithmicend\ \algorithmicswitch}%
\algdef{SE}[CASE]{Case}{EndCase}[1]{\algorithmiccase\ #1}{\algorithmicend\ \algorithmiccase}%
\algtext*{EndSwitch}%
\algtext*{EndCase}%

\def\BibTeX{{\rm B\kern-.05em{\sc i\kern-.025em b}\kern-.08em
    T\kern-.1667em\lower.7ex\hbox{E}\kern-.125emX}}
\usepackage{url}

\usepackage{tikz}
\usetikzlibrary{shapes, arrows, positioning,plotmarks,patterns}

\usepackage[caption=false]{subfig}

\usepackage{pgfplots}
\pgfplotsset{
legend style={fill opacity=0.7,  draw opacity=1, text opacity=1, draw=white!15!black, legend cell align=left, align=left},
width=6cm, 
height=6cm,
yminorticks=false,
xminorticks=false,
title style={font=\small},
tick style={color=black},
tick label style={font=\small},
grid style={line width=.1pt, draw=gray!20},
major grid style={line width=.1pt,draw=gray!20},
}
\usepgfplotslibrary{colorbrewer}
\usepgfplotslibrary{groupplots}
\usepgfplotslibrary{fillbetween}
\usetikzlibrary{patterns}
\pgfplotsset{
compat=1.11,
every tick label/.append style={font=\footnotesize},
legend image code/.code={
\draw[mark repeat=2,mark phase=2]
plot coordinates {
(0cm,0cm)
(0.15cm,0cm)        
(0.3cm,0cm)         
};%
}
}
    
\usepackage{glossaries}
\newacronym{ai}{AI}{Artificial Intelligence}
\newacronym{aoi}{AoI}{Age of Information}
\newacronym{beta}{BETA}{Bandit-based Emergent Threshold Adaptation}
\newacronym{br}{BR}{Best Response}
\newacronym{cdf}{CDF}{Cumulative Distribution Function}
\newacronym{cdns}{cDNS}{canonical DNS}
\newacronym{dns}{DNS}{Dominant Node Strategy}
\newacronym{fbrp}{FBRP}{Finite Best Response Path}
\newacronym{goma}{GoMA}{Goal-oriented Medium Access}
\newacronym{goc}{GoC}{Goal-oriented Communication}
\newacronym{ibr}{IBR}{Iterated Best Response}
\newacronym{iid}{i.i.d.}{independent and identically distributed}
\newacronym{jfi}{JFI}{Jain Fairness Index}
\newacronym{kkt}{KKT}{Karush-Kuhn-Tucker}
\newacronym{libra}{LIBRA}{Local Iterated Best Response Access}
\newacronym{mab}{MAB}{Multi-Armed Bandit}
\newacronym{mac}{MAC}{Medium Access Control}
\newacronym{maf}{MAF}{Maximum Age First}
\newacronym{mc}{MC}{Monte Carlo}
\newacronym{ms}{MS}{Monitoring Station}

\newacronym{ne}{NE}{Nash Equilibrium}
\newacronym{pdf}{PDF}{Probability Density Function}
\newacronym{pmf}{PMF}{Probability Mass Function}
\newacronym{psd}{PSD}{Positive Semidefinite}
\newacronym{voi}{VoI}{Value of Information}
\newacronym{wsn}{WSN}{Wireless Sensor Network}

\DeclareMathOperator*{\Proba}{Pr}    
\newcommand{\E}[1]{\mathbb{E}\left[ #1 \right]} 
\newcommand{\Esub}[2]{\mathbb{E}_{#1}\left[ #2 \right]} 
\newcommand{\Prb}[1]{\Proba\left[ #1 \right]} 
\newcommand{\mc}[1]{\mathcal{#1}}   
\newcommand{\mb}[1]{\mathbf{#1}}    
\DeclareMathOperator*{\argmax}{arg\,max}    
\DeclareMathOperator*{\argmin}{arg\,min}    

\newcounter{Probl}
\newenvironment{problemeq}
  {\stepcounter{Probl}%
    \addtocounter{equation}{-1}%
    \equation}
  {\endequation}

\algtext*{EndWhile}
\algtext*{EndIf}
\algtext*{EndFor}

\def \fwidth{0.99\columnwidth}
\def \fheight {0.42\columnwidth}
\def \twofigw{0.48\linewidth}
\def \twofigh{0.2\linewidth}
\def \threefigw{0.32\linewidth}
\def \threefigh{0.2\linewidth}
\def \boxside{0.3\columnwidth}
\def \boxheight{0.24\columnwidth}

\definecolor{color0}{HTML}{00429D}
\definecolor{color1}{HTML}{844D99}
\definecolor{color2}{HTML}{C3608E}
\definecolor{color3}{HTML}{EF8078}
\definecolor{color14}{HTML}{915a8f}
\definecolor{color34}{HTML}{d27f76}
\definecolor{color4}{HTML}{FFB047}
\definecolor{darkslategray38}{RGB}{38,38,38}

\newtheorem{theorem}{Theorem}

\newtheorem{lemma}{Lemma}[theorem]

\newtheorem{corollary}{Corollary}[theorem]

\makeatletter

\title{A Theory of Goal-Oriented Medium Access: Protocol Design and Distributed Bandit Learning}

\author{
\IEEEauthorblockN{Federico Chiariotti and Andrea Zanella}
\IEEEauthorblockA{Dept. of Information Engineering,
University of Padova, Italy (emails: \{name.lastname\}@unipd.it)}
}

\begin{document}

\maketitle
\begin{abstract}
The \gls{goc} paradigm breaks the separation between communication and the content of the data, tailoring communication decisions to the specific needs of the receiver and targeting application performance. While recent studies show impressive encoding performance in point-to-point scenarios, the multi-node distributed scenario is still almost unexplored. Moreover, the few studies to investigate this consider a centralized collision-free approach, where a central scheduler decides the transmission order of the nodes. In this work, we address the \gls{goma} problem, in which multiple intelligent agents must coordinate to share a wireless channel and avoid mutual interference. We propose a theoretical framework for the analysis and optimization of distributed \gls{goma}, serving as a first step towards its complete characterization. We prove that the problem is non-convex and may admit multiple \gls{ne} solutions. We provide a characterization of each node’s best response to others’ strategies and propose an optimization approach that provably reaches one such \gls{ne}, outperforming centralized approaches by up to $100\%$ while also reducing energy consumption. We also design a distributed learning algorithm that operates with limited feedback and no prior knowledge.
\end{abstract}


\glsresetall

\section{Introduction}\label{sec:intro}

\gls{goc}, first envisioned in Warren Weaver's introduction to Shannon's theory of communication~\cite{shannon1949mathematical}, is a paradigm that considers communication links not as simple bit-pipes, agnostic of the meaning of the carried data, but as a service designed to support the application's goals \cite{gunduz2022beyond}. Although the layering approach that separates the communication problem from the underlying meaning of the data has been a universal success over almost a century, new applications, such as cooperative robotics and autonomous driving, require a deeper integration between communication networks and computational, sensing and control systems~\cite{qi2024architecture}, whose optimization will be deeply intertwined.

The first push towards this integration was the development of \emph{semantic} communication schemes~\cite{bourtsoulatze2019deep}, which integrate source and channel coding through the design of languages that incorporate and exploit the shared context between transmitter and receiver~\cite{shao2024theory}. The recent leap forward in \gls{ai} techniques is enabling the exchange of ever more compact models~\cite{chaccour2024less} along with task-specific data, further improving the adaptability and effectiveness of the semantic communication paradigm. This concept has been expanded to \emph{effective} communication~\cite{kam2024reinforcement}, which aims to assist the receiver directly in performing its task~\cite{diao2025task}, ultimately aiming towards the joint optimization of communication and control~\cite{talli2025pragmatic}.

However, current semantic schemes typically focus on single point-to-point channels with a single decision maker, thus avoiding coordination issues. In other words, these schemes address \textit{what} to transmit, but not \textit{who} should transmit. When considering multi-node scenarios, the coordination of access to shared communication resources based on semantic or effective communication criteria becomes increasingly relevant. The use of tools such as compositional reasoning and dynamic epistemic logic has recently been advocated to allow future networks to truly self-organize~\cite{bennis2025semantic}, but solving the \gls{goma}
problem remains an open and fundamental challenge.

To date, the problem has been mainly addressed in \emph{pull-based} scenarios~\cite{ayan2024optimal}, where a central decision-maker makes use of its statistical knowledge of each node's \gls{voi} to schedule node transmissions orthogonally, avoiding mutual interference~\cite{akar2024query}. Whenever the \gls{voi} of an update is a deterministic function of the \gls{aoi}, the scheduling problem can be optimally solved with index-based policies that compute an optimal ranking of potential transmitters~\cite{tripathi2024whittle}. Instead, when the \gls{voi} is stochastic (e.g., for alarm systems, where some measurements are inherently much more informative than others), channel access can only be optimized in an average sense, as the central decision-maker does not have direct access to the nodes' actual measurements. It should be noted that the practical implementation of pull-based scheduling also requires additional downlink signaling to distribute the schedule~\cite{cavallero2024coexistence}, which may be troublesome in low-power \glspl{wsn}.

In the complementary \textit{push-based} approach~\cite{gunduz2023timely}, the perspective is somehow reversed: each node independently decides whether and when to transmit its data, knowing the \gls{voi} of what was observed and, possibly, the statistical distribution of the \gls{voi} of the other nodes' measurements.  
This approach can potentially reduce unnecessary transmissions, but requires coordination between nodes to avoid mutual interference. 
So far, this coordination has been based on feedback techniques~\cite{madueno2014efficient} and data prioritization~\cite{zhang2023value}, leading to simple heuristics based on two principles: binding transmission or signal power to a threshold criterion on the data \gls{voi}\cite{agheli2024goal}, or using globally available information to estimate other nodes' \gls{voi} based on Bayesian reasoning~\cite{chiariotti2025delta}.

In this work, we propose a theoretical framework for \gls{goma} that applies to both push-based and pull-based approaches and provides an analytical grounding to existing heuristic strategies. We consider a scenario in which cooperative intelligent agents are acquainted with (or estimate) others' \gls{voi} statistical distributions. We then define an optimization problem that aims to determine the transmission strategy of each node that maximizes the expected \gls{voi} at the receiver, while minimizing the total number of transmissions, based on the information available to the nodes. 

Our first theoretical result shows that 
the optimization problem is non-convex, possibly admitting multiple locally optimal solutions. 
Then, following a game-theoretical approach, we show that it is possible to find a locally optimal strategy by applying an \gls{ibr} approach, and we provide a closed-form expression of the optimal response of one node to any combination of strategies of the others. This leads to the \textit{\gls{libra}} protocol, which can find distributed \gls{goma} solutions that significantly outperform any access scheme that relies on a single transmitter dominating, while all the others are almost always silent (as provided by pull-based approaches). Tested in some explanatory scenarios, \gls{libra} is shown to achieve significant gains in terms of average reward over the pull-based like solutions, while also reducing the number of transmission attempts and, thus, the energy consumption of the \gls{wsn}. 
Finally, we relax the assumption about prior knowledge of the \gls{voi} statistical distribution at the nodes, and propose the \textit{\gls{beta}}. This learning algorithm, rooted in multi-agent semi-bandits~\cite{kale2010non}, is shown to converge quasi-exponentially to the \gls{libra} solution without any prior information on \gls{voi} statistics.

The rest of this paper is organized as follows. Sec.~\ref{sec:system} presents the system model with a toy example showing the potential gains of push-based approaches. We then pose the \gls{goma} problem in Sec.~\ref{sec:solution}. Sec.~\ref{sec:libra} describes the idea behind \gls{libra} and its implementation, while \gls{beta} is presented in Sec.~\ref{sec:bandit}. A behavior and performance analysis of the two solutions is presented in Sec.~\ref{sec:results}, and finally, Sec.~\ref{sec:conc} draws our conclusions and presents possible avenues for future work.

\section{System Model}\label{sec:system}

Consider a scenario in which a \gls{ms} observes a random process through measurements collected by a set $\mc{N}$ of heterogeneous sensor nodes. The \gls{voi} $v_n$ of node $n$'s measurement is modeled as a discrete random variable $V_n$ taking values from set $\mathcal{V}_n$, following a distribution with \gls{pmf} $p_n$.\footnote{For simplicity and without any loss of accuracy, in the following we assume that the \gls{voi} is directly \textit{measured} by nodes, rather than determined from the observation of the environment. Moreover, we assume $p_{n,v}>0$ for all $v\in\mc{V}$, since values that are never observed are of no interest. } We assume that the \glspl{voi} of the nodes are mutually independent, and that each node can determine the actual value $v_n$ of its measurement, but it only knows the distribution $p_m$ of the \gls{voi} of any other node $m$. The nodes share an ideal time-slotted collision channel: if a single node transmits, its packet is always decoded correctly, while if multiple nodes transmit, all packets are erased.\footnote{This model can be extended to more complex channels with wireless errors or capture probabilities. However, this would complicate the notation and analysis without changing the fundamental properties of the system.
} We also consider a fixed transmission cost $\psi$, which can represent the energy expenditure of the transmission.

\subsection{Pull-Based (Dominant Node) Access Model}
In pull-based communication, the \gls{ms} polls a single node for transmission, avoiding collisions. The index of the polled node corresponds to the action $a^{\text{pull}}\in \mathcal{N}$ taken by the \gls{ms}. The polled node transmits an update only if its \gls{voi} is higher than the transmission cost $\psi$, so 
 the reward for the \gls{ms} is  
 \begin{equation}\label{eq:rpull}
R^{\text{pull}}(a^{\text{pull}})=\max\{0,V_{a^{\text{pull}}}-\psi\}\,.
 \end{equation}
Unfortunately, the \gls{ms} cannot predict \eqref{eq:rpull}, because it does not know the \gls{voi} of the nodes in advance. Therefore, the optimal action is to poll the node with the maximum \textit{weighted tail expectation} of $V_n$ with respect to the threshold $\psi$, i.e., 
\begin{equation}\label{eq:pull_action}
a^{\text{pull}}=\argmax_{n\in\mathcal{N}}\E{V_n|V_n>\psi}\Prb{V_n>\psi}.
\end{equation}
Note that with this approach, the \gls{ms} always polls the node that offers the maximum expected reward. 

In the following, any joint strategy where the transmission probability of one node is much higher than that of the others is named \textit{\gls{dns}}. A joint strategy is said to be a \emph{\gls{cdns}} when the transmission probability is $1$ for dominant node and $0$ for the others. 

\subsection{Push-Based Access Model}
A more complex model is required for push-based strategies, where each node acts based on its own value $v_n$ and on the knowledge of the statistical distribution of every other node's \gls{voi}. We denote the strategy of node $n$ as $\mb{x}_n\in[0,1]^{|\mc{V}_n|}$, where each element $x_{n,v}$ is the transmission probability of $n$ when it observes value $v$. We denote the expected transmission probability of node $n$ as:
\begin{equation}
    \bar{x}_n=\Esub{v\sim p_n}{x_{n,v}}=\sum_{v\in\mc{V}_n}p_{n,v}x_{n,v}.
\end{equation}
In addition, we introduce the auxiliary term 
\begin{equation}
    \zeta_n=\prod_{\ell\neq n}(1-\bar{x}_{\ell}),
\end{equation}
which gives the probability that all nodes other than $n$ remain silent. 
The overall transmission strategy, $\mb{x}$, is the collection of the transmission strategies of employed by all nodes. The expected reward obtained by $\mb{x}$ can be expressed as 
\begin{equation}\label{eq:discrete_reward}\E{R|\mb{x}}=\sum_{n\in\mc{N}}\sum_{v\in\mc{V}_n}v\zeta_np_{n,v}x_{n,v}-\psi\sum_{n\in\mc{N}}\bar{x}_n\,,
\end{equation}
where the first term on the left-hand side is the expected \gls{voi} of successful transmissions, whereas the other term represents the average energy cost of all transmission attempts.
Our goal is then to solve the following constrained minimization problem:
\begin{problemeq}\label{prob:discrete}
\begin{aligned}
    \mb{x}^*=&\argmin_{\mb{x}\in\mathbb{R}^{|\mc{N}|\times|\mc{V}|}}-\E{R|\mb{x}},\\
    \text{such that }&\mb{x}_n\in[0,1]^{|\mc{V}_n|}\ \forall n\in\mc{N}.
\end{aligned}
\end{problemeq}

\begin{figure}
\subfloat[$\psi=0$. \label{fig:bin_psi0}]{\begin{tikzpicture}
\begin{axis}[
    width=\boxside,
    height=\boxheight,
    scale only axis,
    xlabel={$p$},
    ylabel={$\E{R}$},
    xmin=0, xmax=1,
    ymin=0, ymax=1,
    legend pos=south east,
    ymajorgrids=true,
    xmajorgrids=true,
    grid style=dashed,
    legend style={legend columns=1, font=\tiny, anchor=south east, at={(1,0)}}
]

\addplot[
    color=color0,
    mark=x, mark repeat=10, mark phase=0
    ]
    table[x=p,y=Vpull] {fig/fig_data/binary_psi0.dat};
\addlegendentry{DNS};
\addplot[
    color=color1,
    mark=o, mark repeat=10, mark phase=8
    ]
    table[x=p,y=V2] {fig/fig_data/binary_psi0.dat};
\addlegendentry{$N\!=\!2$};
\addplot[
    color=color2,
    mark=triangle, mark repeat=10, mark phase=6
    ]
    table[x=p,y=V5] {fig/fig_data/binary_psi0.dat};
\addlegendentry{$N\!=\!5$};
\addplot[
    color=color3,
    mark=diamond, mark repeat=10, mark phase=4
    ]
    table[x=p,y=V10] {fig/fig_data/binary_psi0.dat};
\addlegendentry{$N\!=\!10$};
\addplot[
    color=color4,
    mark=+, mark repeat=10, mark phase=2
    ]
    table[x=p,y=V100] {fig/fig_data/binary_psi0.dat};
\addlegendentry{$N\!=\!100$};
\end{axis}
\end{tikzpicture}}
\subfloat[$\psi=0.25$.\label{fig:bin_psi025}]{\begin{tikzpicture}
\begin{axis}[
    width=\boxside,
    height=\boxheight,
    scale only axis,
    xlabel={$p$},
    ylabel={$\E{R}$},
    xmin=0, xmax=1,
    ymin=0, ymax=1,
    legend pos=south east,
    ymajorgrids=true,
    xmajorgrids=true,
    grid style=dashed,
    legend style={legend columns=1, font=\tiny, anchor=south east, at={(0.98,0.02)}}
]

\addplot[
    color=color0,
    mark=x, mark repeat=10, mark phase=0
    ]
    table[x=p,y=Vpull] {fig/fig_data/binary_psi025.dat};
\addplot[
    color=color1,
    mark=o, mark repeat=10, mark phase=8
    ]
    table[x=p,y=V2] {fig/fig_data/binary_psi025.dat};
\addplot[
    color=color2,
    mark=triangle, mark repeat=10, mark phase=6
    ]
    table[x=p,y=V5] {fig/fig_data/binary_psi025.dat};
\addplot[
    color=color3,
    mark=diamond, mark repeat=10, mark phase=4
    ]
    table[x=p,y=V10] {fig/fig_data/binary_psi025.dat};
\addplot[
    color=color4,
    mark=+, mark repeat=10, mark phase=2
    ]
    table[x=p,y=V100] {fig/fig_data/binary_psi025.dat};
\end{axis}
\end{tikzpicture}}
\caption{Push- and pull-based communication performance under a binary symmetric value distribution.}\vspace{-0.4cm}
\label{fig:binary_example}
\end{figure}
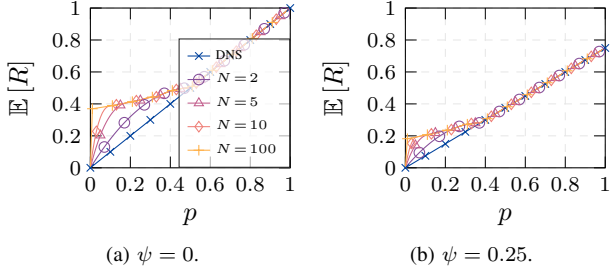

\subsection{Toy Example}
To appreciate the importance of finding push-based solutions, rather than settle for a \gls{dns}, we consider a simple motivating example. We assume that any node $n\in \mathcal{N}$ observes anomalies ($v\!=\!1$) with probability $p_n(1)\!=\!p$, or a normal state ($v=0$) with $p_n(0)\!=\!1\!-\!p$, independently of the others. The \gls{ms}'s goal is to be notified about observed anomalies. 

In a pull-based approach, one single node is required to transmit if it observes an anomaly, while the others are always silent. As the nodes are statistically identical, all \glspl{dns} are equivalent. In a push-based system, instead, more nodes are allowed to transmit with probability $x_{n,1}\geq 0$ if they observe an anomaly. In this simple case, $\mb{x}^*$ can be found in closed form, but we omit the derivation due to space constraints. 

The performance of these two schemes for zero and positive transmission costs is shown in Fig.~\ref{fig:bin_psi0} and Fig.~\ref{fig:bin_psi025}, respectively, when varying the probability of anomaly $p$ and the number $N$ of nodes (which only affects the push-based approach). We can see that the \gls{dns} is always outperformed by the push-based optimal scheme. The performance gap is higher when $p$ is low, that is, if anomalies at any individual node are rare, while the gap vanishes for 
$p>0.5$, as the push-based solution essentially converges to a \gls{dns}, in which all nodes but one are always silent. 
Moreover, we note that the performance gain of the push-based method in the low-$p$ region increases with the number of nodes, but decreases for a positive transmission cost $\psi$, since the overhead caused by collisions becomes heavier.

This example shows that a distributed, push-based approach can outperform the best possible \gls{dns}, motivating the need for a formal analysis of the push-based \gls{goma} problem in more complex scenarios, with larger value sets $\mc{V}$ and asymmetric statistical distributions of the \gls{voi}. 

\section{Mathematical Modeling of GoMA}\label{sec:solution}
In this section, we focus on the constrained optimization problem \eqref{prob:discrete}. For simplicity, we assume that $\mc{V}$ is the same for all nodes, but our results can be directly extended to the general case where nodes have different value domains. We start from the following observation. 

\begin{lemma}\label{lm:discrete_nonconvexity}
    Problem~\eqref{prob:discrete} is non-convex.
\end{lemma}
\begin{IEEEproof}
To study the (non)convexity of \eqref{prob:discrete}, we perform a definiteness analysis over the Hessian matrix $\nabla^2$ of the cost function, which is the additive inverse of the expected reward 
in~\eqref{eq:discrete_reward}. The element of $\nabla^2$ with respect to $x_{n,v}$ and $x_{m,u}$ is
\begin{equation}
    \nabla^2_{n,v,m,u}=(1-\delta_{m,n})
        (u+v)\zeta_np_{n,v}p_{m,u}(1-\bar{x}_m)^{-1},
\end{equation}
where $\delta_{m,n}$ is $1$ if $m=n$ and $0$ otherwise. 
We note that $\nabla^2$ is a hollow real matrix, as it is symmetric with all diagonal elements equal to $0$. Sylvester's criterion for a matrix to be \gls{psd} is that all its principal minors must be \gls{psd}. At least one $2\times 2$ minor $\mb{M}$ of $\nabla^2$ is structured as
\begin{equation}
    \mb{M}=\begin{bmatrix}
    0 & (u+v)p_{n,v}p_{m,u}\\
    (u+v)p_{n,v}p_{m,u} & 0
    \end{bmatrix}.
\end{equation}
This minor is \gls{psd} if and only if
\begin{equation}
    \mb{z}^T\mb{M}\mb{z}=2z_1z_2(u+v)p_{n,v}p_{m,u} \geq 0\ \forall\mb{z}\neq0.
\end{equation}
However, setting $z_1=-z_2$, and considering $n$, $v$, $m$, and $u$ for which $p_{n,v}\neq 0$ and $p_{m,u}\neq 0$, the result is negative. Hence, the Hessian is not \gls{psd} and the problem is non-convex.
\end{IEEEproof}
Therefore, the problem may admit multiple locally optimal solutions. 
We then consider another perspective, posing the \gls{goma} optimization as a game-theoretical problem. Each node is a player that must decide whether or not to transmit, given its current observed \gls{voi} and \textit{knowing the transmission strategies of the other players}. The reward function for each player is the expected \gls{voi} of successful transmissions minus the transmission cost. Due to the purely cooperative nature of the problem, we can exploit the properties of potential games~\cite{monderer1996potential} to reach an \gls{ne}, which is a local optimum, through \gls{br} dynamics~\cite{young1993evolution}.
\begin{theorem}\label{th:nash_discrete}
  Adopting an \acrfull{ibr} approach for each player leads to an $\varepsilon$-\gls{ne}, i.e., a solution with an expected reward within $\varepsilon$ from that of an \gls{ne}, after a number of steps that grows polynomially with $1/\varepsilon$.
\end{theorem}
\begin{IEEEproof}
  As all players have the same reward function, the game is an exact potential game. Using \gls{ibr} over players then leads to a \gls{fbrp}~\cite[Thm.~3]{milchtaich1996congestion}: each \gls{br} increases the potential function, leading to an \gls{ne} of the game, in a finite number of steps. The \gls{br} dynamics converge to an $\varepsilon$-\gls{ne} in polynomial time~\cite{sun2023provably}.
\end{IEEEproof}

This result allows us to solve~\eqref{prob:discrete} iteratively by determining the \gls{br} of each node, as explained in the following. 

\subsection{The Best Response Problem}
The \gls{fbrp} resulting from the iterative procedure will always lead to an $\varepsilon$-\gls{ne}, but finding such a path requires solving the \emph{\acrfull{br}} problem. In this problem, we fix $\bm{x}_{-n}$, i.e., the strategy of all nodes except node $n$, and find the strategy that maximizes the reward:
\begin{problemeq}\label{prob:discrete_br}
\begin{aligned}
    \mb{x}^*_n(\bm{x}_{-n})=&\argmax_{\mb{x}_n\in\mathbb{R}^{|\mc{V}|}}\E{R|\mb{x}_n,\bm{x}_{-n}},\\
    \text{such that }&\mb{x}_n\in[0,1]^V.
\end{aligned}
\end{problemeq}
The following result states that at least one \gls{br} of a player to the transmission strategies of the others is threshold-based. 

\begin{theorem}\label{th:best_response_threshold}
    For every node $n$ and every strategy $\bm{x}_{-n}$, one optimal solution $\mb{x}^*_n$ of problem~\eqref{prob:discrete_br} is a threshold strategy such that  $x^*_{n,v}(\bm{x}_{-n})=0,\, \forall v\leq\theta_n^*(\bm{x}_{-n})$ and $x^*_{n,v}(\bm{x}_{-n})=1,\, \forall v>\theta_n^*(\bm{x}_{-n})$, with a threshold $\theta_n^*(\bm{x}_{-n})$ given by \begin{equation}\label{eq:theta_discrete}
    \theta_n^*(\bm{x}_{-n})=\sum_{m\neq n}\sum_{u\in\mc{V}}\frac{up_{m,u}x_{m,u}}{(1-\bar{x}_m)}+\frac{\psi}{\zeta_n}\,.
    \end{equation}
\end{theorem}
 \begin{IEEEproof}
We first observe that, for any solution $\mb{x}_n$, we can partition the value set $\mc{V}$ into three subsets in such a way that $\mc{A}_0$ collects all and only the measurements $v\in\mc{V}$ that are never transmitted ($x_{n,v}=0$), $\mc{A}_1$ collects those that are always transmitted ($x_{n,v}=1$), and $\mc{B}$ those transmitted with a positive but not certain probability ($0<x_{n,v}<1$).

Then, we note that Problem~\eqref{prob:discrete_br} is linear, as the inequality constraints are affine functions and, once $\bm{x}_{-n}$ is fixed, the reward  in~\eqref{eq:discrete_reward} also becomes an affine function of $\mb{x}_n$. Therefore, it can be solved using the \gls{kkt} conditions~\cite{kuhn1951nonlinear}. We hence consider the Lagrange function:
\begin{equation}\label{eq:lagrange}
    \mc{L}=\E{R|\mb{x}_n,\bm{x}_{-n}}+\sum_{v\in\mc{V}}\lambda_{n,v}x_{n,v}-\lambda'_{n,v}(1-x_{n,v})\,,
\end{equation}
where $\lambda_{n,v}$ and $\lambda'_{n,v}$ are non-negative multipliers associated to the problem's constraints. 
The \gls{kkt} theorem states that, for a point $\mb{x}_n$ to be optimal, it must satisfy the stationarity and complementary slackness conditions. The stationarity condition requires that the gradient of $\mc{L}$ with respect to $\mb{x}_{n}$ must be zero at the optimal point. Then, setting to zero the partial derivative of \eqref{eq:lagrange} with respect to $x_{n,v}$, we get 
\begin{equation}\label{eq:lagrangian_derivative}
\zeta_n p_{n,v}(v- \theta_n^*)= \lambda'_{n,v}-\lambda_{n,v}\,,
\end{equation}
with $ \theta_n^*$ as given in \eqref{eq:theta_discrete} (we omit the argument to reduce clutter).
The complementary slackness conditions, on the other hand, require that $\lambda_{n,v}x_{n,v}=0$ and $\lambda'_{n,v}(1-x_{n,v})=0$, for all $v\in\mc{V}$. Applying these conditions to \eqref{eq:lagrangian_derivative} for the values $v\in \mc{A}_0$, we get $
\lambda_{n,v}=-p_{n,v}\zeta_n(v\!- \theta_n^*)
$ and, since $\lambda_{n,v}$ must be non-negative, we have 
$v \leq \theta_n^*$. Similarly, for all $v\in\mc{A}_1$ we get $
\lambda_{n,v}'=p_{n,v}\zeta_n(v\!-\theta_n^*)
$ form which $v \geq\theta_n^*$. Finally, for $v\in\mc{B}$, it must be
$v = \theta_n^* \in \mc{V}$, and set $\mc{B}$ is empty otherwise. Hence, the three subsets of $\mc{V}$ corresponds to a partition based on the threshold $\theta_n^*$. 

Note that, if $ \theta_n^* \notin \mc{V}$, the subset $\mc{B}$ is empty. In this case, $\mb{x}_n$ is a purely threshold-based strategy. Conversely, if $\theta_n \in \mc{V}$, we need to prove that there exists at least one purely threshold-based strategy that is equally optimal. 
Now, if $\theta_n^* \in \mc{V}$, using \eqref{eq:theta_discrete} in the reward expression \eqref{eq:discrete_reward} we obtain:
\begin{equation}
\begin{aligned}
\E{R|\theta_n^*,x_{n,\theta^*_n},\bm{x}_{-n}}=\sum_{v>\theta^*_n}p_{n,v}\left[v\zeta_n-\psi\right]-\sum_{m\neq n}\bar{x}_m\psi\\
+\sum_{m\neq n}\sum_{v\in\mc{V}}u \frac{\zeta_{n}p_{m,v}x_{m,v}}{1-\bar{x}_m}\left[1-\sum_{v>\theta^*_n}p_{n,v}\right]\,,
\end{aligned}
\end{equation}
which does not depend on $x_{n,\theta_n}$. Therefore, all strategies $\mb{x}'_n$ such that $x'_{n,v}=x_{n,v}$ for all $v\in \mc{V}\setminus \{\theta^*_n\}$ are equally optimal, included the purely threshold-based one with $x'_{n,\theta_n}=0$. 
\end{IEEEproof}

\begin{corollary}\label{cor:threshold}
   We can define a threshold-based problem  whose optimum is also a locally optimal solution of~\eqref{prob:discrete}:
   \begin{problemeq}\label{prob:discrete_thresh}
\begin{aligned}
    \bm{\theta}^*_n=&\argmin_{\bm{\theta}\in\mathbb{\mc{V}}^{|\mc{N}|}}-\E{R|\bm{\theta}}.
\end{aligned}
\end{problemeq}
\end{corollary}
\begin{IEEEproof}
    The corollary trivially follows from Theorem~\ref{th:best_response_threshold}, as any optimal solution of~\eqref{prob:discrete} must also be optimal for~\eqref{prob:discrete_br} for every node $n$. The optimal threshold-based strategy is then also a locally optimal strategy of the general problem.
\end{IEEEproof}
Finding the solution to problem~\eqref{prob:discrete_thresh} directly requires a time that is at most polynomial in $|\mc{V}|$, as the feasible set has $|\mc{V}|^{|\mc{N}|}$ elements. However, the complexity quickly grows with the number of agents, and brute force approaches become unfeasible for large networks or value spaces, requiring the lighter \gls{ibr} approach. If all nodes apply a threshold strategy (collectively denoted as $\bm{\theta}_{-n}$), the \gls{br} threshold in~\eqref{eq:theta_discrete} becomes
\begin{equation}\label{eq:theta_thresh_discrete}
    \theta_n^*(\bm{\theta}_{-n})=\sum_{m\neq n}\frac{\sum_{u>\theta_m}up_{m,u}}{1-\sum_{{w>\theta_m}}p_{m,w}}+\frac{\psi}{\zeta_n}.
\end{equation}

\subsection{Extension to Continuous Value Domains}

Let us consider an extension of the model, in which $\mc{V}$ is a convex subset of $\mathbb{R}^+$. In this case, the value distribution is defined by its \gls{cdf} $P_n$, and the transmission strategy $\mb{x}_n$ becomes a function of $v$. We limit ourselves to the case in which the expected \gls{voi} is finite, i.e., $\E{V_n}<\infty$.
We use the probability integral transform to get $x_n:[0,1]\to[0,1]$, associating a transmission probability to each quantile through function $Q_n:[0,1]\to\mc{V}$, defined as $Q_n(p)=\inf\left\{v\in{V}:P_n(v)\geq p\right\}$.
The quantile function is monotonically increasing. The  total transmission probability becomes $\bar{x}_n=\int_0^1x_{n}(p)dp$. We can then rewrite~\eqref{eq:discrete_reward} as
\begin{equation}\label{eq:continuous_reward}
    \E{R|\mb{x}}=\sum_{n\in\mc{N}}\left[\zeta_n\int_0^1Q_n(p)x_{n}(p)dp-\psi\bar{x}_n\right].
\end{equation}
We can define the \gls{br} problem in this case as
\begin{problemeq}\label{prob:cont_br}
    \mb{x}^*_n(\bm{x}_{-n})=\argmin_{\mb{x}_n\in L_2(0,1)}-\E{R|\mb{x}_n,\bm{x}_{-n}},
\end{problemeq}
where $L_2(0,1)$ is the Hilbert space containing all functions from $[0,1]$ to $[0,1]$.
\begin{theorem}\label{th:cont_best_response_threshold}
At least one optimal solution $\mb{x}^*_n(\bm{x}_{-n})$ of problem~\eqref{prob:cont_br} is a threshold strategy for which $x^*_{n,v}(\bm{x}_{-n})=0,\, \forall v<\theta_n^*(\bm{x}_{-n})$ and $x^*_{n,v}(\bm{x}_{-n})=1,\, \forall v\geq\theta_n^*(\bm{x}_{-n})$ for any node $n$ and any strategy $\bm{x}_{-n}$, where $\theta_n^*(\bm{x}_{-n})$ is
\begin{equation}
\label{eq:tyhresholdcont}
  \theta_n^*(\bm{x}_{-n})=P_n\left(\frac{\psi}{\theta_n}+\sum_{\mathclap{m\neq n}} \frac{\int_{0}^1Q_n(p)x_m(p)dp}{1-\bar{x}_m}\right).
\end{equation}   
\end{theorem}
\begin{IEEEproof}
The Hilbert space $L_2(0,1)=x:[0,1]\to[0,1]$ is convex and closed. As integrals are linear functionals~\cite{goldstein1964convex}, the \gls{kkt} conditions can be extended~\cite{russell1966kuhn}. We can show that there is an optimal threshold strategy by following the proof of Theorem~\ref{th:best_response_threshold}. 
We then obtain the \gls{br} $\theta_n^*(\bm{x}_{-n})$ by computing the derivative of the expected reward with respect to $\theta_n$:
\begin{equation}
\frac{\partial\E{R|\theta_n,\bm{x}_{-n}}}{\partial\theta_n}=\zeta_n\!\left[\sum_{\mathclap{\ell\neq n}} \int_{\mathclap{\ \ \ \theta_{\ell}}}^{1}\frac{x_{\ell}(p)Q_{\ell}(p)}{1-\bar{x}_{\ell}}dp\!-\!Q_n(\theta_n)\!\right]\!+\psi.\label{eq:der1}
\end{equation}
The derivative is equal to $0$, excluding the trivial cases in which other nodes have $\bar{x}_m=1$, if and only if
\begin{equation}
  \theta_n^*(\bm{x}_{-n})=P_n\left(\frac{\psi}{\zeta_n}+\sum_{\mathclap{m\neq n}} \frac{\int_{0}^1Q_m(p)x_m(p)dp}{1-\bar{x}_m}\right).
  \label{eq:threshold2}
\end{equation}
If the quantile function is not strictly increasing, \eqref{eq:threshold2} may admit multiple solutions. In this case, we can easily prove that any such value is optimal, and we arbitrarily consider the infimum of the set, as per our definition of the quantile function.
The derivative in \eqref{eq:der1} is positive if $\theta_n<\theta_n^*$ and negative if $\theta_n>\theta_n^*$, as all elements of the sum are positive. As such, the point is the minimum of problem~\eqref{prob:cont_br}.
\end{IEEEproof}

Corollary~\ref{cor:threshold} can also be extended to the continuous value case, and the \gls{br} function for threshold strategies is
\begin{equation}
   \theta_n^*(\bm{\theta}_{-n})=P_n\left(\frac{\psi}{\zeta_n}+\sum_{\mathclap{m\neq n}} \frac{\int_{\theta_m}^1Q_m(p)dp}{\theta_m}\right).   
\end{equation}

\section{The LIBRA Algorithm}\label{sec:libra}

Theorem~\ref{th:nash_discrete} allows us to reach an $\varepsilon$-\gls{ne} by starting from an initial strategy $\bm{\theta}^{(0)}$ and iteratively computing the \gls{br} for each node until convergence. However, the problem may admit multiple \glspl{ne}. For example, if $\psi=0$, all the $|\mc{N}|$ \glspl{cdns} where a single node always transmits are \glspl{ne}, since any unilateral deviation from this strategy would reduce the average reward. 
We then need to find the initial strategy that leads to the highest-reward \gls{ne}, i.e., the global optimum.
 To gain more insight, we analyze the \glspl{ne} in an explanatory use case. 

\subsection{Attraction Regions of \glspl{ne}}

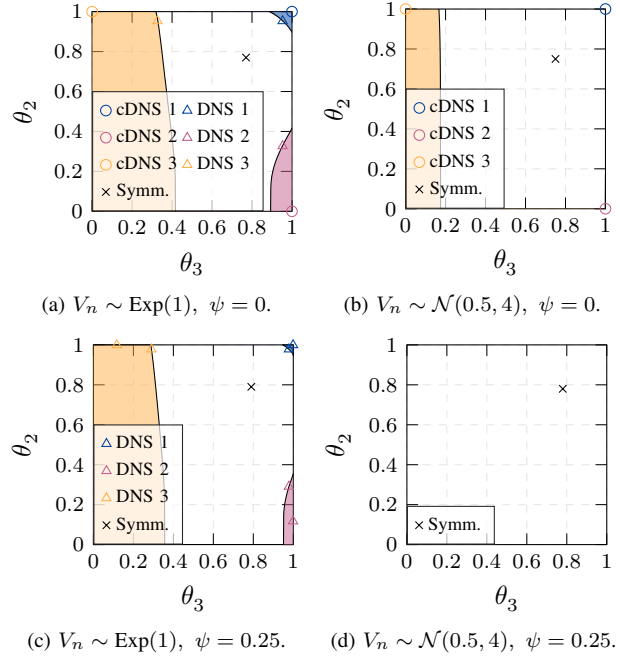
\begin{figure}
\centering
\subfloat[$V_n\sim\text{Exp}(1),\ \psi=0$. \label{fig:attr_exp}]{\begin{tikzpicture}
\begin{axis}[
    width=\boxside,
    height=\boxside,
    scale only axis,
    xlabel={$\theta_3$},
    ylabel={$\theta_2$},
    xmin=0, xmax=1,
    ymin=0, ymax=1,
    legend pos=south east,
    ymajorgrids=true,
    xmajorgrids=true,
    grid style=dashed,
    legend style={font=\scriptsize, anchor=south west,legend columns=2, at={(0,0)}}
]

\path[name path=axis_low] (axis cs:0,0) -- (axis cs:1,0);
\path[name path=axis_high] (axis cs:0,1) -- (axis cs:1,1);
\addplot[color=black, name path=theta1_a, forget plot] 
    table[x=x1,y=x2] {fig/fig_data/attractors_exp_theta1_a.dat};
\addplot[
    color=black, name path=theta1_b, forget plot]
    table[x=x1,y=x2] {fig/fig_data/attractors_exp_theta1_b.dat};

\addplot[
    color=black, name path=theta1_c, forget plot]
    table[x=x1,y=x2] {fig/fig_data/attractors_exp_theta1_c.dat};

\addplot[
        fill=color0, 
        fill opacity=0.5, forget plot
    ]
    fill between[
        of=theta1_b and axis_high,
        soft clip={domain=0:1},
    ];

\addplot[
        fill=color4, 
        fill opacity=0.5, forget plot
    ]
    fill between[
        of=theta1_a and axis_low,
        soft clip={domain=0:1},
    ];

\addplot[
        fill=color2, 
        fill opacity=0.5, forget plot
    ]
    fill between[
        of=theta1_c and axis_low,
        soft clip={domain=0:1},
    ];

\addplot[only marks, mark=o, color=color0]
table{%
  1 1
};
  \addlegendentry{cDNS 1};

\addplot[only marks, mark=triangle, color=color0]
table{%
  0.953 0.953
};
  \addlegendentry{DNS 1};

\addplot[only marks, mark=o, color=color2]
table{%
  1 0
};
  \addlegendentry{cDNS 2};

\addplot[only marks, mark=triangle, color=color2]
table{%
  0.953 0.327
};
  \addlegendentry{DNS 2};

\addplot[only marks, mark=o, color=color4]
table{%
  0 1
};
  \addlegendentry{cDNS 3};

\addplot[only marks, mark=triangle, color=color4]
table{%
  0.327  0.953
};
  \addlegendentry{DNS 3};

  \addplot[only marks, mark=x, color=black]
table{%
  0.77  0.77
};
  \addlegendentry{Symm.};


\end{axis}
\end{tikzpicture}}
\subfloat[$V_n\sim\mc{N}(0.5,4),\ \psi=0$.\label{fig:attr_gauss}]{\begin{tikzpicture}
\begin{axis}[
    width=\boxside,
    height=\boxside,
    scale only axis,
    xlabel={$\theta^{}_3$},
    ylabel={$\theta^{}_2$},
    xmin=0, xmax=1,
    ymin=0, ymax=1,
    legend pos=south east,
    ymajorgrids=true,
    xmajorgrids=true,
    grid style=dashed,
    legend style={font=\scriptsize, anchor=south west, at={(0,0)}}
]

\path[name path=axis_low] (axis cs:0,0) -- (axis cs:1,0);
\path[name path=axis_high] (axis cs:0,1) -- (axis cs:1,1);
\addplot[color=black, name path=theta1_a, forget plot] 
    table[x=x1,y=x2] {fig/fig_data/attractors_gauss_theta1_a.dat};
\addplot[
    color=black, name path=theta1_b, forget plot]
    table[x=x1,y=x2] {fig/fig_data/attractors_gauss_theta1_b.dat};

\addplot[
        fill=color4, 
        fill opacity=0.5, forget plot
    ]
    fill between[
        of=theta1_b and axis_low,
        soft clip={domain=0:1},
    ];

\addplot[
        fill=color4, 
        fill opacity=0.5, forget plot
    ]
    fill between[
        of=theta1_a and axis_low,
        soft clip={domain=0:1},
    ];

\addplot[only marks, mark=o, color=color0]
table{%
  1 1
};
  \addlegendentry{cDNS 1};

\addplot[only marks, mark=o, color=color2]
table{%
  1 0
};
  \addlegendentry{cDNS 2};

\addplot[only marks, mark=o, color=color4]
table{%
  0 1
};
  \addlegendentry{cDNS 3};

\addplot[only marks, mark=x, color=black]
table{%
  0.75  0.75
};
  \addlegendentry{Symm.};


\end{axis}
\end{tikzpicture}}\\ \vspace{-0.2cm}
\subfloat[$V_n\sim\text{Exp}(1),\ \psi=0.25$. \label{fig:attr_exp_psi}]{\begin{tikzpicture}
\begin{axis}[
    width=\boxside,
    height=\boxside,
    scale only axis,
    xlabel={$\theta^{}_3$},
    ylabel={$\theta^{}_2$},
    xmin=0, xmax=1,
    ymin=0, ymax=1,
    legend pos=south east,
    ymajorgrids=true,
    xmajorgrids=true,
    grid style=dashed,
    legend style={font=\scriptsize, anchor=south west, at={(0,0)}}
]

\path[name path=axis_low] (axis cs:0,0) -- (axis cs:1,0);
\path[name path=axis_high] (axis cs:0,1) -- (axis cs:1,1);
\addplot[color=black, name path=theta1_a, forget plot] 
    table[x=x1,y=x2] {fig/fig_data/attractors_exp025_theta1_a.dat};
\addplot[
    color=black, name path=theta1_b, forget plot]
    table[x=x1,y=x2] {fig/fig_data/attractors_exp025_theta1_b.dat};

\addplot[
    color=black, name path=theta1_c, forget plot]
    table[x=x1,y=x2] {fig/fig_data/attractors_exp025_theta1_c.dat};

\addplot[
        fill=color0, 
        fill opacity=0.5, forget plot
    ]
    fill between[
        of=theta1_b and axis_high,
        soft clip={domain=0:1},
    ];

\addplot[
        fill=color4, 
        fill opacity=0.5, forget plot
    ]
    fill between[
        of=theta1_a and axis_low,
        soft clip={domain=0:1},
    ];

\addplot[
        fill=color2, 
        fill opacity=0.5, forget plot
    ]
    fill between[
        of=theta1_c and axis_low,
        soft clip={domain=0:1},
    ];

\addplot[only marks, mark=triangle, color=color0]
table{%
  0.999 0.999
  0.978 0.978
};
  \addlegendentry{DNS 1};

\addplot[only marks, mark=triangle, color=color2]
table{%
  0.999 0.117
  0.978 0.2905
};
  \addlegendentry{DNS 2};

  \addplot[only marks, mark=triangle, color=color4]
table{%
  0.117 0.999
  0.2905  0.978
};
  \addlegendentry{DNS 3};

\addplot[only marks, mark=x, color=black]
table{%
  0.79  0.79
};
  \addlegendentry{Symm.};


\end{axis}
\end{tikzpicture}}
\subfloat[$V_n\sim\mc{N}(0.5,4),\ \psi=0.25$.\label{fig:attr_gauss_psi}]{\begin{tikzpicture}
\begin{axis}[
    width=\boxside,
    height=\boxside,
    scale only axis,
    xlabel={$\theta^{}_3$},
    ylabel={$\theta^{}_2$},
    xmin=0, xmax=1,
    ymin=0, ymax=1,
    legend pos=south east,
    ymajorgrids=true,
    xmajorgrids=true,
    grid style=dashed,
    legend style={font=\scriptsize, anchor=south west, at={(0,0)}}
]

\path[name path=axis_low] (axis cs:0,0) -- (axis cs:1,0);
\path[name path=axis_high] (axis cs:0,1) -- (axis cs:1,1);

\addplot[only marks, mark=x, color=black]
table{%
  0.78  0.78
};
  \addlegendentry{Symm.};


\end{axis}
\end{tikzpicture}}

\caption{Attraction regions of various \glspl{ne} in a symmetric scenario with $3$ nodes, \gls{iid} \glspl{voi} with Exponential (left-hand side) and Normal (right-hand side) distribution, and zero (upper) and positive (lower) transmission cost. }\vspace{-0.4cm}
\label{fig:attractors}
\end{figure}

We examine a simple scenario with $3$ nodes, each employing a threshold-based transmission strategy. As node $1$ immediately adjusts its threshold to match the other two nodes, the initial conditions are solely determined by $\theta^{(0)}_2$ and $\theta^{(0)}_3$.\footnote{We recall that, with continuous \gls{voi}, the \gls{br} thresholds are given by \eqref{eq:tyhresholdcont}.} We divide this space into a grid and apply \gls{ibr} from each starting point until convergence. Fig.~\ref{fig:attractors} shows the resulting \glspl{ne} (markers), under two \gls{voi} distributions (Exponential and Gaussian) and energy cost terms ($\psi=0$ and $\psi=0.25$). 

In all cases, there exists a symmetric \gls{ne} where all three nodes transmit with equal probability (cross marker). However, the number and location of other \glspl{ne} vary depending on the specific settings. As anticipated, when the energy cost coefficient is zero ($\psi=0$) the three \gls{cdns} solutions emerge as \glspl{ne} (circle markers at the corners of Fig.~\ref{fig:attr_exp}-\subref*{fig:attr_gauss}). The white area corresponds to initial solutions that converge to the symmetric \gls{ne}, while the colored areas represent the attraction basins of the \glspl{cdns}. We note that almost all initial point on the left-hand side of the figures (low $\theta_3^{(0)}$) lead to the \gls{cdns} at the top-left corner, in which node $3$ dominates. In fact, nodes $1$ and $2$ adapt to the relatively high transmission probability of node $3$ by decreasing their own transmission probabilities to avoid collisions. In turn, this allows node $3$ to further reduce its threshold at the end of the \gls{ibr} round. The process eventually converges to \gls{cdns} dominated by node $3$. The other two nodes also have \glspl{cdns}, with much smaller attraction regions. In addition, we see three more \gls{dns} solutions (triangle markers), placed at the border of the attraction regions of the symmetric and \gls{cdns} solutions. These other \glspl{dns} 
correspond to cases where one node transmits if it observes $v_n>0.4$, while the other two transmit only if their \gls{voi} is exceptionally large, i.e., $v_m>3$. They have a lower reward than all other solutions, with negligible attraction basins.
Finally, the reward gap between the symmetric push-based \gls{ne} and the \glspl{cdns} is approximately $1\%$ under exponential distribution of \gls{voi}, and $10\%$ in the Gaussian case, accounting for the different sizes of the attraction regions of the \glspl{cdns} shown in Fig.~\ref{fig:attr_exp}-\subref*{fig:attr_gauss}.

Considering the case with $\psi=0.25$, shown in Fig.~\ref{fig:attr_exp_psi}-\subref*{fig:attr_gauss_psi}, we observe that the attraction regions of the \glspl{cdns} shrink significantly and, in the Gaussian case, all \gls{dns} \glspl{ne} disappear.
Fig.~\ref{fig:attr_exp_psi} shows that the \glspl{ne} closest to the upper left corner is no longer a \gls{cdns}, since node 3 transmits with probability lower than $1$ ($\theta_3>0$). Accordingly, we mark this solution using a triangle marker, like for the other non-canonical \glspl{dns}.
The other two nodes transmit if their observed value is in the extreme right tail, hence with probability $0.001$. The same can be observed for the other nodes, both of which dominate in two non-canonical \glspl{dns}, although with smaller attraction regions, for the reasons described above.
In the Gaussian case (Fig.~\ref{fig:attr_gauss_psi}), the symmetric solution becomes the only \gls{ne}. This counterintuitive effect arises because, with $\psi=0.25$, the transmission of lower values would yield a negative reward. Therefore, solutions where a single node transmits almost independently of the observed value become suboptimal, and the \gls{br} dynamics eventually leads to the symmetric equilibrium.

\subsection{The LIBRA Algorithm}

From the above analysis, it emerges that, in symmetric scenarios, \glspl{ne} close to the center of the feasible set have relatively large attraction basins and
are often preferable to \gls{dns} solutions, providing more balanced strategies. We then present the \gls{libra} scheme, which applies \gls{ibr} from a cleverly chosen initial strategy. The core idea, in fact, is to choose a starting strategy that can lead to a non-\gls{dns} \gls{ne}.

We then restrict our choice of  the  starting point to the set of so-called \emph{equal value} strategies, for which
\begin{equation}Q_m(\theta^{(0)}_m)=Q_n(\theta^{(0)}_n),\ \forall m,n\in\mc{N}.
\end{equation}
Intuitively, these strategies allow the nodes to share the channel fairly, but they may not be optimal even if the nodes' values are \gls{iid}: as we learned from the binary scenario in Fig.~\ref{fig:binary_example}, for $p>\frac{1}{N}$, equal value threshold strategies are always beaten by strategies in which some nodes never transmit. Nonetheless, an equal value strategy can be good for \gls{ibr} initialization.

\begin{table}[b]
 \vspace{-0.3cm}
    \centering
    \caption{Equal value strategy outcomes in \gls{iid} conditions.}
    \begin{tabular}{l|cc}
    \toprule
        Distribution & $\E{V_n|V_n>v_{\text{eq}}}$ & $v_{\text{eq}}\ (\psi=0)$ \\
        \midrule
        Uniform ($0,V$) & $\frac{V+v_{\text{eq}}}{2}$ & $V\sqrt{\frac{N-1}{N+1}}$ \\
         Exponential ($\lambda$) & $\frac{1}{\lambda}+v_{\text{eq}}$ & --- \\
        Gaussian ($\mu,\sigma^2$) & $\mu+\frac{\sigma e^{-\frac{(v_{\text{eq}}-\mu)^2}{2\sigma^2}}}{\sqrt{2\pi}(1-P_n(v_{\text{eq}}))}$ & ---\\
        Pareto ($z,\alpha)$ & $\frac{\alpha v_{\text{eq}}}{\alpha-1}$ & $z\left(\frac{\alpha(N-1)}{\alpha N - 1}\right)^{-\frac{1}{\alpha}}$\\
    \bottomrule
    \end{tabular}
    \label{tab:symmetrical_solutions}
\end{table}

We then want to find the \textit{optimal} equal value strategy to start our iterative method. In the discrete case, this is possible by iterating over all possible values $v\in\mc{V}$ and finding the one that maximizes the reward when set as the transmission threshold, while in the continuous case, the value space can be quantized with a given precision. Starting from the expected reward function in~\eqref{eq:continuous_reward}, we rewrite the expected reward as
\begin{equation}
    \E{R|v_{\text{eq}}}=\!\sum_{n\in\mc{N}}\int_{\mathclap{\quad\quad\ P_n(v_{\text{eq}})}}^1\!Q_n(p)dp \prod_{\mathclap{m\neq n}} P_m(v_{\text{eq}})-\psi\!\sum_{\mathclap{n\in\mc{N}}}(1-P_n(v_{\text{eq}})).
\end{equation}
Optimizing $v_{\text{eq}}$ will then depend on the distribution and, in some cases, may only be feasible through an exhaustive search. However, since it is a single parameter, the complexity of getting within a distance $\varphi$ of the optimum is $O\left(\varphi^{-1}\right)$. The values of $\E{V_n|V_n>Q_n(\theta)}$ for some common distributions under \gls{iid} nodes are given in Table~\ref{tab:symmetrical_solutions}, along with the closed-form solution where it exists.

The pseudocode for \gls{libra} is given in Alg.~\ref{alg:ibr}. The complexity of \gls{ibr} is $O\left(N\varepsilon^{-1}\right)$, as it requires $O\left(\varepsilon^{-1}\right)$ iterations~\cite{sun2023provably} and each \gls{br} round is a closed-form expression that can be computed in $O(N)$ time. If we set $\varphi=\varepsilon$, the initialization also requires $O\left(\varepsilon^{-1}\right)$ operations under exhaustive search.

\begin{figure}[t]
\vspace{-8pt}
\centering
\begin{algorithm}[H]
\caption{The \gls{libra} threshold computation scheme}
\label{alg:ibr}
\begin{algorithmic}[1]
\scriptsize

\Require $\mc{N},\mc{V},\mb{P},\psi$
\State $\bm{\theta}^{(0)}\gets$\Call{EqualValueInitialization}{$\mc{N},\mc{V},\mb{P},\psi$}
\State $\bm{\theta}^{(1)}\gets\mb{0}$, $i\gets 1$
\While {$\bm{\theta}^{(i)}\neq\bm{\theta}^{(i-1)}$}
    \State $\bm{\theta}^{(i)}\gets\bm{\theta}^{(i-1)}$
    \For{$n\in\mc{N}$}
        \State $\theta^{(i)}_n\gets$\Call{BestResponse}{$\mc{N},\mc{V},\mb{P},\psi,\bm{\theta}^{(i)}_{-n}$}\Comment{Update using~\eqref{eq:theta_thresh_discrete}}
    \EndFor
    \State $i\gets i+1$
\EndWhile
\State\Return{$\bm{\theta}^{(i)}$}
\end{algorithmic}
\end{algorithm}
 \vspace{-0.6cm}
\end{figure}

\section{Emergent Goal-Oriented Medium Access}\label{sec:bandit}
The previous analysis is based on the assumption that all nodes have full knowledge of the \gls{voi} distribution of all other nodes. In this section, we relax this assumption and propose the \textit{\acrfull{beta}}, an approach based on \glspl{mab} that does not require prior knowledge of the value distributions.

We hence assume that the nodes have no initial knowledge of their own or any other node's \gls{voi} distribution. After each transmission slot, the receiver broadcasts a feedback with the \gls{voi} of the received update, if any, or the idle or collided state of the slot, otherwise. The nodes periodically report to the receiver the fraction of slots $\rho_n$ in which they attempted to transmit. The receiver then broadcasts the aggregate transmission rate $\bar{\rho}$ of all nodes as additional feedback.

Thanks to Corollary~\ref{cor:threshold}, we limit our problem to threshold strategies: each node maintains a \gls{mab} whose arms are the possible values of $\theta_n$. Naturally, rewards will be stochastic, as they depend on the actual realization of the value for each node as well as the selected thresholds. The reward sample for the selected action, $r^{(i)}(\theta_n(i);\bm{\theta}_{-n})$, is a noisy observation of the expected reward:\begin{equation}
    r^{(i)}(\theta_n(i);\bm{\theta}_{-n}(i))=\E{R(\theta_n(i);\bm{\theta}_{-n}(i))}+\xi_n(i),
\end{equation}
where $\xi_n(i)$ is a martingale noise process that represents the stochasticity of the \gls{voi} distributions~\cite{cohen2017learning}. 

However, feedback allows us to consider the problem as a \emph{semi-bandit} scenario, in which each node is able to obtain not only a reward sample for the selected action, $r_n^{(i)}(\theta_n(i);\bm{\theta}_{-n})$, but also an estimate $\hat{r}_n^{(i)}(\theta;\bm{\theta}_{-n})$ for any other possible choice of $\theta_n(i)$. These rewards samples are collected in vector $\hat{\mb{r}}_n^{(i)}$.

\subsection{The $\varepsilon$-Hedge Learning Algorithm}
The well-known $\varepsilon$-Hedge algorithm~\cite{freund1997decision} has an exploration rate $\varepsilon\in(0,1)$. During exploration slots, which occur with probability $\varepsilon$, the threshold is chosen following a uniform random distribution, while in other slots $\varepsilon$-Hedge uses reward estimates $\bar{\mb{r}}_n^{(i-1)}$ as the weights of a softmax distribution:
\begin{equation}
    P(\theta_n^{(i)}=\theta)=\frac{e^{\bar{r}_n^{(i-1)}(\theta)}}{\sum_{\theta'\in\mc{V}}e^{\bar{r}_n^{(i-1)}(\theta')}}.
\end{equation}
The initial reward estimates $\bar{\mb{r}}_n^{(0)}$ need to be given as an input. The update rule for semi-bandit feedback $\hat{\mb{r}}^{(i)}_n$ is
\begin{equation}
    \bar{\mb{r}}_n^{(i)}= \bar{\mb{r}}_n^{(i-1)}+\gamma_i\hat{\mb{r}}^{(i)}_n,
\end{equation}
where $\bm{\gamma}\propto i^{-\kappa},\ \kappa\in\left(\frac{1}{2},1\right)$ is a decreasing step-size sequence.

\begin{figure}[t]
\vspace{-8pt}
\centering

\begin{algorithm}[H]
\caption{Semi-bandit feedback estimation}
\label{alg:semibandit}
\begin{algorithmic}[1]
\scriptsize

\Require $\mc{V}$, $\psi$, $\theta_n(i)$, $\omega$, $V_n(i)$, $v(i)$, $\bar{\rho}(i)$, $\rho_{n}(i)$, $\alpha_n(i)$, $\beta_n(i)$, $\lambda_n(i)$

  \For{$\theta\in\mc{V}$}
  \State $a\gets (\theta< v_n(i))$\Comment{True if node $n$ would have transmitted}
  \Switch{$\omega$}
    \Case{Silence}
      \State $\hat{r}_n^{(i)}(\theta) \gets a(v_n(i)-\psi)$
    \EndCase
    \Case{Success}
      \If{$\theta_n(i)<v_n(i)$} \Comment{The update was from $n$}
        \State $\hat{r}_n^{(i)}(\theta) \gets a(v_n(i)-\psi)$ 
      \Else \Comment{The update was from another node}
        \State $\hat{r}_n^{(i)}(\theta) \gets (1-a)v(i)-(1+a)\psi$
      \EndIf
    \EndCase
    \Case{Collision}
      \If{$\theta_n(i)<v_n(i)$}
            \Comment{Node $n$ collided}
        \State $\hat{r}_n^{(i)}(\theta)\gets (1-a)\lambda_n(i)-\frac{\psi(\bar{\rho}(i)-\rho_n(i))}{\alpha_n(i)}-\psi a$
      \Else \Comment{Node $n$ was silent}
        \State $\hat{r}_n^{(i)}(\theta) \gets -\frac{\psi(\bar{\rho}(i)-\rho_n(i)-\beta_n(i))}{\alpha_n(i)-\beta_n(i)}-\psi a$
    \EndIf
    \EndCase
    \EndSwitch
  \EndFor
  \State \Return{$\hat{\mb{r}}_n^{(i)}$}
\end{algorithmic}
\end{algorithm}
\vspace{-0.5cm}
\end{figure}

As proved in \cite{cohen2017learning}, the $\varepsilon$-Hedge algorithm converges quasi-exponentially to the optimal joint solution  with probability $1$, although its worst-case regret is $O\left(n^{-1/2}\right)$, when $\hat{\mb{r}}^{(i)}_n$ is a finite-variance, unbiased estimator of $\E{R}$. In the next section, we prove that a semi-bandit feedback function $\hat{\mb{r}}^{(i)}_n$ with these properties can be obtained by each node $n$ by performing counterfactual reasoning. Under these mild conditions, nodes are hence able to quickly converge to an $\varepsilon$-\gls{ne} solution.

\subsection{Counterfactual Semi-Bandit Reward Estimation}
We then consider the semi-bandit reward estimation from the perspective of node $n$, given the reported outcome $\omega$.

\begin{figure*}
\centering
\subfloat[Reward. \label{fig:sens_rew}]{\begin{tikzpicture}
\begin{axis}[
    width=\twofigw,
    height=\twofigh,
    xlabel={$N$},
    ylabel={$\E{R}$},
    xmin=0, xmax=100,
    ymin=0, ymax=2.2,
    legend pos=south east,
    ymajorgrids=true,
    xmajorgrids=true,
    grid style=dashed,
    legend style={legend columns=2, font=\scriptsize, anchor=south east, at={(1,0)}}
]

\addplot[
    color=color0
    ]
    table[x=N,y=pull] {fig/fig_data/sensors_symm_rewards_psi0.dat};
\addlegendentry{cDNS ($\psi\!=\!0$)};
\addplot[
    color=color0,     dashed
    ]
    table[x=N,y=pull] {fig/fig_data/sensors_symm_rewards_psi25.dat};
\addlegendentry{cDNS ($\psi\!=\!0.25$)};

\addplot[
    color=color2
    ]
    table[x=N,y=th] {fig/fig_data/sensors_symm_rewards_psi0.dat};
\addlegendentry{LIBRA (th., $\psi\!=\!0$)};
\addplot[
    color=color2, only marks,
    mark=triangle, mark repeat=10, mark phase=6
    ]
    table[x=N,y=mc] {fig/fig_data/sensors_symm_rewards_psi0.dat};
\addlegendentry{LIBRA (MC, $\psi\!=\!0$)};

\addplot[
    color=color2,dashed
    ]
    table[x=N,y=th] {fig/fig_data/sensors_symm_rewards_psi25.dat};
\addlegendentry{LIBRA (th., $\psi\!=\!0.25$)};
\addplot[
    color=color2, only marks,
    mark=x, mark repeat=10, mark phase=1
    ]
    table[x=N,y=mc] {fig/fig_data/sensors_symm_rewards_psi25.dat};
\addlegendentry{LIBRA (MC, $\psi\!=\!0.25$)};

\end{axis}
\end{tikzpicture}}
\subfloat[Total energy consumption. \label{fig:sens_en}]{\begin{tikzpicture}
\begin{axis}[
    width=\twofigw,
    height=\twofigh,
    xlabel={$N$},
    ylabel={$E$},
    xmin=0, xmax=100,
    ymin=0, ymax=1.05,
    legend pos=south east,
    ymajorgrids=true,
    xmajorgrids=true,
    grid style=dashed,
    legend style={legend columns=2, font=\scriptsize, anchor=south east, at={(1,0)}}
]

\addplot[
    color=color0
    ]
    table[x=N,y=pull] {fig/fig_data/sensors_symm_energy_psi0.dat};
\addlegendentry{cDNS ($\psi\!=\!0$)};
\addplot[
    color=color0,     dashed
    ]
    table[x=N,y=pull] {fig/fig_data/sensors_symm_energy_psi25.dat};
\addlegendentry{cDNS ($\psi\!=\!0.25$)};

\addplot[
    color=color2
    ]
    table[x=N,y=th] {fig/fig_data/sensors_symm_energy_psi0.dat};
\addlegendentry{LIBRA (th., $\psi\!=\!0$)};
\addplot[
    color=color2, only marks,
    mark=triangle, mark repeat=10, mark phase=6
    ]
    table[x=N,y=mc] {fig/fig_data/sensors_symm_energy_psi0.dat};
\addlegendentry{LIBRA (MC, $\psi\!=\!0$)};

\addplot[
    color=color2,dashed
    ]
    table[x=N,y=th] {fig/fig_data/sensors_symm_energy_psi25.dat};
\addlegendentry{LIBRA (th., $\psi\!=\!0.25$)};
\addplot[
    color=color2, only marks,
    mark=x, mark repeat=10, mark phase=1
    ]
    table[x=N,y=mc] {fig/fig_data/sensors_symm_energy_psi25.dat};
\addlegendentry{LIBRA (MC, $\psi\!=\!0.25$)};

\end{axis}
\end{tikzpicture}}
\caption{\gls{libra} performance in the symmetric \gls{wsn} scenario.}\vspace{-0.5cm}
\label{fig:wsn_symm}
\end{figure*}
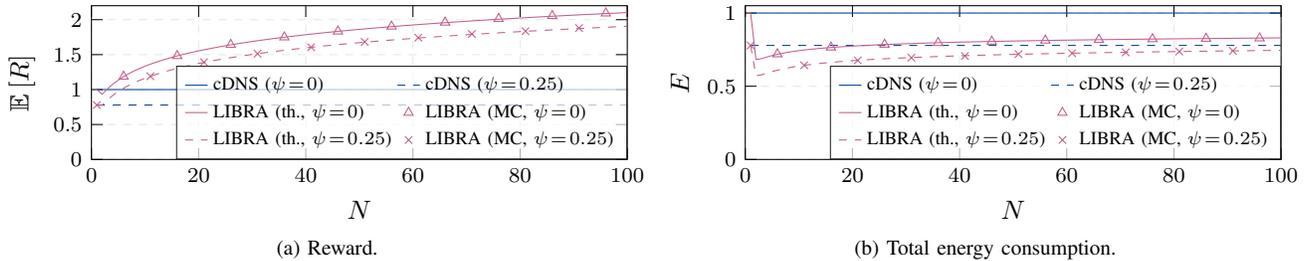

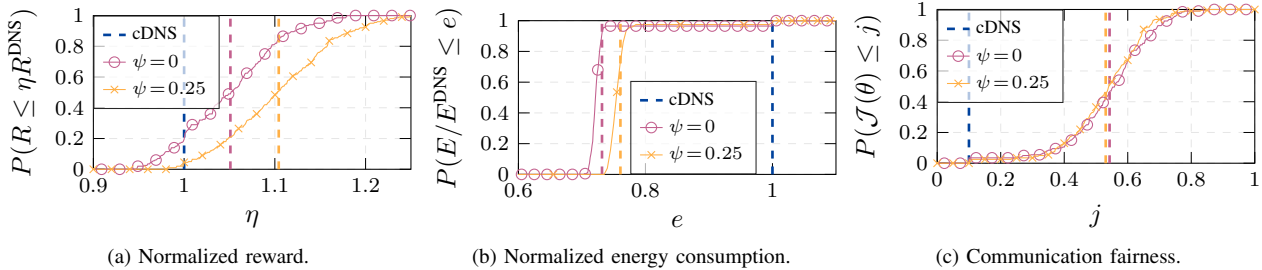
\begin{figure*}
\centering
\subfloat[Normalized reward. \label{fig:sens_asymm_gain}]{\begin{tikzpicture}
\begin{axis}[
    width=\threefigw,
    height=\threefigh,
    xlabel={$\eta$},
    ylabel={$P(R\leq\eta R^{\text{DNS}})$},
    xmin=0.9, xmax=1.25,
    ymin=0, ymax=1,
    legend pos=south east,
    ymajorgrids=true,
    xmajorgrids=true,
    grid style=dashed,
    legend style={legend columns=1, font=\scriptsize, anchor=north west, at={(0,1)}}
]

\addplot[
    color=color0, line width=1pt, dashed
    ]
    table {
1  0
1  1
    };
\addlegendentry{cDNS};

\addplot[
    color=color2, mark=o, mark repeat=20, mark phase=10
    ]
    table[x=gain,y=cdf] {fig/fig_data/sensors_asymm_gain_psi0.dat};
\addlegendentry{$\psi\!=\!0$};
\addplot[
    color=color2, line width=1pt, dashed, forget plot
    ]
    table {
1.0510  0
1.0510  1
    };
\addplot[
    color=color4, mark=x, mark repeat=20
    ]
    table[x=gain,y=cdf] {fig/fig_data/sensors_asymm_gain_psi25.dat};
\addlegendentry{$\psi\!=\!0.25$};

\addplot[
    color=color4, line width=1pt, dashed, forget plot
    ]
    table {
1.1043  0
1.1043  1
    };

\end{axis}
\end{tikzpicture}}
\subfloat[Normalized energy consumption. \label{fig:sens_asymm_en}]{\begin{tikzpicture}
\begin{axis}[
    width=\threefigw,
    height=\threefigh,
    xlabel={$e$},
    ylabel={$P(E/E^{\text{DNS}}\leq e)$},
    xmin=0.6, xmax=1.1,
    ymin=0, ymax=1,
    legend pos=south east,
    ymajorgrids=true,
    xmajorgrids=true,
    grid style=dashed,
    legend style={legend columns=1, font=\scriptsize, anchor=south east, at={(0.75,0)}}
]

\addplot[
    color=color0, line width=1pt, dashed
    ]
    table {
1 0
1 1
    };
\addlegendentry{cDNS};

\addplot[
    color=color2, mark=o, mark repeat=20, mark phase=6
    ]
    table[x=rel_en,y=cdf] {fig/fig_data/sensors_asymm_energy_psi0.dat};
\addlegendentry{$\psi\!=\!0$};

\addplot[
    color=color2, line width=1pt, dashed, forget plot
    ]
    table {
0.7316  0
0.7316  1
    };

\addplot[
    color=color4, mark=x, mark repeat=20, mark phase=12
    ]
    table[x=rel_en,y=cdf] {fig/fig_data/sensors_asymm_energy_psi25.dat};
\addlegendentry{$\psi\!=\!0.25$};

\addplot[
    color=color4, line width=1pt, dashed, forget plot
    ]
    table {
0.7606  0
0.7606  1
    };

\end{axis}
\end{tikzpicture}}
\subfloat[Communication fairness. \label{fig:sens_asymm_fair}]{\begin{tikzpicture}
\begin{axis}[
    width=\threefigw,
    height=\threefigh,
    xlabel={$j$},
    ylabel={$P(\mc{J}(\mb{\theta})\leq j)$},
    xmin=0, xmax=1,
    ymin=0, ymax=1,
    legend pos=south east,
    ymajorgrids=true,
    xmajorgrids=true,
    grid style=dashed,
    legend style={legend columns=1, font=\scriptsize, anchor=north west, at={(0,1)}}
]

\addplot[
    color=color0, line width=1pt, dashed
    ]
    table {
0.1 0
0.1 1
    };
\addlegendentry{cDNS};

\addplot[
    color=color2, mark=o, mark repeat=20, mark phase=10
    ]
    table[x=jfi,y=cdf] {fig/fig_data/sensors_asymm_jfi_psi0.dat};
\addlegendentry{$\psi\!=\!0$};

\addplot[
    color=color2, line width=1pt, dashed, forget plot
    ]
    table {
0.5428  0
0.5428  1
    };

\addplot[
    color=color4, mark=x, mark repeat=20
    ]
    table[x=jfi,y=cdf] {fig/fig_data/sensors_asymm_jfi_psi25.dat};
\addlegendentry{$\psi\!=\!0.25$};

\addplot[
    color=color4, line width=1pt, dashed, forget plot
    ]
    table {
0.5306  0
0.5306  1
    };

\end{axis}
\end{tikzpicture}}
\caption{\gls{libra} performance over $200$ episodes in the asymmetric \gls{wsn} scenario with $N=10$ and $\nu=0.5$, relative to the push-based solution. The average gain for each setting is reported as a vertical dashed line with the same color.}
\label{fig:wsn_asymm}
\end{figure*}

\begin{theorem}\label{th:bandit}
If all \gls{voi} distributions have a finite variance, 
there exists a finite-variance, unbiased estimator of $\E{R}$.
\end{theorem}
\begin{IEEEproof}
We first suppose that node $n$ does not transmit at a certain slot and consider the possible outcomes $\omega$. 
If  $\omega$ indicates an idle slot, the reward estimation is 
$\hat{r}_n^{(i)}(\theta)=v_n(i)-\psi$ for any $\theta\leq v_n(i)$, as node $n$ would transmit (while all other nodes are silent). Instead, the reward estimate is zero for any $\theta>v_n(i)$, as node $n$ would have avoided transmission and the slot would have remained idle. 
If $\omega$ indicates a successful transmission, then another node $m\neq n$ was the only one transmitting. Consequently, the reward estimation $\hat{r}_n^{(i)}(\theta)=-2\psi$ for $\theta\leq v_n(i)$, as node $n$ would have transmitted, colliding with node $m$. Conversely, $\hat{r}_n^{(i)}(\theta)=v_m(i)-\psi$ for  $\theta> v_n(i)$, as node $n$ would be silent, enabling node $m$ to successfully complete its transmission. Note that $v_m(i)$ is known by all nodes, being part of the receiver's feedback.

On the other hand, if $\omega$ indicates a collision, then the expected number of colliding nodes, given that at least two have transmitted (not counting $n$), can be estimated as 
$$
N_c = (\bar{\rho}(i) -\rho_n(i)-\beta_n(i))\alpha_n(i)^{-1},
$$
where $\rho_n(i)$ is the overall average number of nodes  that transmit in a slot, $\rho_n(i)$ is the average transmission probability of node $n$, $\beta_n(i)$ is the average number of successful transmissions in the slots where $n$ is silent, and $\alpha_n(i)$ is the fraction of slots in which node $n$ observes transmissions from at least one other node transmitted, regardless of the outcome. All these values can be derived from the receiver feedback or measured by the node itself. 
The reward estimation is then 
$\hat{r}_n^{(i)}(\theta) = -(N_c+1)\psi$ for $\theta\leq v_n(i)$ (since node $n$ would transmit as well, increasing the number of colliders by one), and 
$\hat{r}_n^{(i)}(\theta) = -N_c\psi$ for $\theta > v_n(i)$. 

Let us now consider the case in which node $n$ transmitted in slot $i$, and derive the reward estimations for the possible outcomes. In this case, the slot cannot be idle, as we know that node $n$ transmitted.
In the event $\omega$ reported a successful slot, which means that node $n$ was the only transmitter, the reward estimation is the same as for the case of an idle outcome. The case of a collision outcome, finally, is slightly more involved, since the reward depends on how many other nodes transmitted in the same slot. Let $N_c^\prime = (\bar{\rho}(i) -\rho_n(i))\alpha_n(i)^{-1}$ indicate the mean number of nodes other than $n$ transmitting in a slot, given that at least one of these nodes transmits. Now, for any $\theta\leq v_n(i)$, node $n$ would have transmitted and experienced a collision with at least another node. The reward estimation in this case is $\hat{r}_n^{(i)}(\theta) = -\psi (N^\prime_c+1)$. Instead, for $\theta > v_n(i)$, node $n$ would have refrained from transmission. The outcome and the reward in this case depend on the number of other nodes that have transmitted, on the condition that at least one node has transmitted. The reward can then be estimated as $\hat{r}_n^{(i)}(\theta) = \lambda_n(i)-\psi N_c^\prime$, where $\lambda_n(i)$ is the average value of the transmissions in the slots where node $n$ was silent, given that at least another node transmitted. Naturally, this term is $0$ when the outcome is a collision, and $v_m(i)$ when another node $m$ transmits. This  term can be determined by node $n$ from the receiver's feedback. The reward is then obtained by subtracting the average energy cost of collisions from $\lambda_n(i)$, considering only $N_c'$ colliders due to the silence of node $n$. 

All these
estimates are unbiased and with finite variance, provided that the strategies of nodes are stable. There is in fact a trade-off between the accuracy of the estimates and the risk of bias, as using a longer window to compute the running averages might include samples with outdated strategies, but also reduces the variance. 
In any case, the semi-bandit feedback respects the conditions
for quasi-exponential convergence.
\end{IEEEproof}

The semi-bandit feedback counterfactual reasoning is given as Alg.~\ref{alg:semibandit}, and has a very low computational complexity.

 \begin{table}[b]
 \vspace{-0.3cm}
    \centering
    \caption{Main simulation parameters.}
    \begin{tabular}{cc|c|c}
    \toprule
    
    \multicolumn{2}{c}{\textbf{Parameter}} & \textbf{Meaning} & \textbf{Value}\\
\midrule
    \multirow{4}{*}{Scenario} & $\psi$ & Energy cost & $\{0,0.25\}$\\
    & $N$ & Number of nodes & $10$\\
    & $T$ & Monte Carlo duration & $10^6$~steps\\
    & $\iota$ & \gls{voi} granularity & $10^{-3}$\\
    & $\nu$ & \gls{voi} variation & $\{0.25,0.5\}$\\
    \midrule
   \multirow{5}{*}{\gls{beta}} & $\varepsilon$ & $\varepsilon$-Hedge exploration rate & $0.01$\\
   & $\kappa$ & Learning rate decay & $1-5\cdot10^{-5}$\\
   & $L$ & Training duration & $10^5$~steps\\
   & $W$ & Traffic estimation window & $25$~steps\\
   & $\mc{V}$ & Possible threshold \glspl{voi} & $\{0,0.1,\ldots,20\}$\\
    \bottomrule
    \end{tabular}
    \label{tab:params}
\end{table}

\section{Simulation Settings and Results}\label{sec:results}

This section presents the evaluation of \gls{beta} and \gls{libra}, which was performed by comparing the analytical results to a Monte Carlo simulation over $10^6$ steps. The main parameters of the simulations and the settings of the \gls{beta} protocol are listed in Table~\ref{tab:params}. The performance of the two protocols is evaluated against the best possible \gls{cdns}, which can be found using state-of-the-art pull-based methods.

In all cases, the Monte Carlo results match our theoretical analysis, confirming its validity.\footnote{The full simulation code is available at \url{https://github.com/signetlabdei/goal_oriented_medium_access}.}
The key metric we use in our evaluation is the expected reward $\E{R}$ as defined in~\eqref{eq:discrete_reward}, but we also consider the average number of transmissions per slot $E=\sum_{n=1}^N\bar{x}_n$, which is a proxy of the energy cost and is crucial in \glspl{wsn}, and the \gls{jfi} $\mc{J}(\bm{\theta})$ over the transmission rate of each node, defined as:
\begin{equation}
    \mc{J}(\bm{\theta})=\frac{\left(\sum_{n\in\mc{N}}(1-\theta_n)\right)^2}{N\sum_{n\in\mc{N}}(1-\theta_n)^2}.
\end{equation}

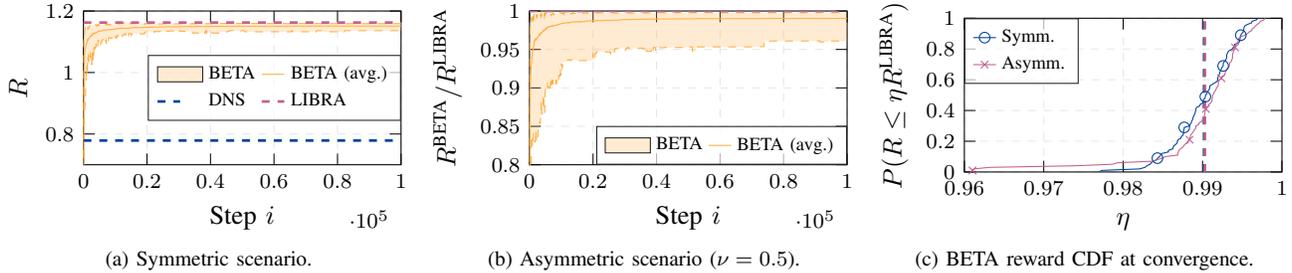
\begin{figure*}
\centering
\subfloat[Symmetric scenario. \label{fig:mab_symm}]{\begin{tikzpicture}
\begin{axis}[
    width=\threefigw,
    height=\threefigh,
    xlabel={Step $i$},
    ylabel={$R$},
    xmin=0, xmax=100000,
    ymin=0.7, ymax=1.2,
    legend pos=south east,
    ymajorgrids=true,
    xmajorgrids=true,
    grid style=dashed,
    legend style={legend columns=2, font=\scriptsize, anchor=east, at={(1,0.5)}}
]

\addplot[
    color=color4, dashed, forget plot, name path=mab_min
    ]
    table[x=step,y=min] {fig/fig_data/mab_symm_conv.dat};

\addplot[
    color=color4, dashed, forget plot, name path=mab_max
    ]
    table[x=step,y=max] {fig/fig_data/mab_symm_conv.dat};

\addplot[
        fill=color4, 
        fill opacity=0.25
    ]
    fill between[
        of=mab_min and mab_max
    ];
\addlegendentry{BETA};

\addplot[
    color=color4
    ]
    table[x=step,y=avg] {fig/fig_data/mab_symm_conv.dat};
\addlegendentry{BETA (avg.)};

\addplot[
    color=color0, line width=1pt, dashed
    ]
    table {
0   0.7788
100000  0.7788
    };
\addlegendentry{DNS};

\addplot[
    color=color2, line width=1pt, dashed
    ]
    table{
0   1.1625
100000  1.1625
};
\addlegendentry{LIBRA};

\end{axis}
\end{tikzpicture}}
\subfloat[Asymmetric scenario ($\nu=0.5$). \label{fig:mab_asymm}]{\begin{tikzpicture}
\begin{axis}[
    width=\threefigw,
    height=\threefigh,
    xlabel={Step $i$},
    ylabel={$R^{\text{BETA}}/R^{\text{LIBRA}}$},
    xmin=0, xmax=100000,
    ymin=0.8, ymax=1,
    legend pos=south east,
    ymajorgrids=true,
    xmajorgrids=true,
    grid style=dashed,
    legend style={legend columns=2, font=\scriptsize, anchor=south east, at={(1,0)}}
]

\addplot[
    color=color4, dashed, forget plot, name path=mab_min
    ]
    table[x=step,y=min] {fig/fig_data/mab_asymm_conv.dat};

\addplot[
    color=color4, dashed, forget plot, name path=mab_max
    ]
    table[x=step,y=max] {fig/fig_data/mab_asymm_conv.dat};

\addplot[
        fill=color4, 
        fill opacity=0.25
    ]
    fill between[
        of=mab_min and mab_max
    ];
\addlegendentry{BETA};

\addplot[
    color=color4
    ]
    table[x=step,y=avg] {fig/fig_data/mab_asymm_conv.dat};
\addlegendentry{BETA (avg.)};

\addplot[
    color=color2, line width=1pt, dashed
    ]
    table{
0   1
100000  1
};

\end{axis}
\end{tikzpicture}}
\subfloat[\gls{beta} reward \gls{cdf} at convergence. \label{fig:mab_final}]{\begin{tikzpicture}
\begin{axis}[
    width=\threefigw,
    height=\threefigh,
    xlabel={$\eta$},
    ylabel={$P(R\leq \eta R^{\text{LIBRA}})$},
    xmin=0.96, xmax=1,
    ymin=0, ymax=1,
    legend pos=south east,
    ymajorgrids=true,
    xmajorgrids=true,
    grid style=dashed,
    legend style={legend columns=1, font=\scriptsize, anchor=north west, at={(0,1)}}
]

\addplot[
    color=color0, mark=o, mark repeat=20, mark phase=10
    ]
    table[x=norm_rew,y=cdf] {fig/fig_data/mab_symm_final.dat};
\addlegendentry{Symm.};
\addplot[
    color=color0, line width=1pt, dashed, forget plot
    ]
    table {
0.9902  0
0.9902  1
    };
\addplot[
    color=color2, mark=x, mark repeat=20
    ]
    table[x=norm_rew,y=cdf] {fig/fig_data/mab_asymm_final.dat};
\addlegendentry{Asymm.};

\addplot[
    color=color2, line width=1pt, dashed, forget plot
    ]
    table {
0.9903  0
0.9903  1
    };

\end{axis}
\end{tikzpicture}}
\caption{\gls{beta} performance over $100$ runs in both scenarios with $N=10$, $\psi=0.25$. The shaded area represents the achievable convergence region.}
\label{fig:mab_conv}\vspace{-0.3cm}
\end{figure*}

In the following, we consider a \gls{wsn} scenario, in which sensors measure independent values and transmit them to a central \gls{ms}, which maintains a Kalman filter~\cite{kalman1960new} or a more general non-linear filter over a Wiener process~\cite{stratonovich1959optimal}. In this case, the innovation from each transmission will be a zero-mean Gaussian random variable, and its square will be a $\chi^2$ random variable with $2$ degrees of freedom. The square of the innovation is a good proxy for the usefulness of a measurement in such filters, although the long-term effect of scheduling decisions is beyond the scope of this work. Consequently, we consider $N$ sensors whose \glspl{voi} follow this model.

We first consider a symmetric scenario, in which all nodes have the same expected \gls{voi} (the distribution is normalized so that $\E{V_n}=1$). Fig.~\ref{fig:sens_rew} shows the comparison between \gls{libra} and the pull-based solution in terms of the expected reward: we note that, whenever $N>2$, \gls{libra} obtains a significantly higher reward, gaining over $50\%$ for $N=20$ and $100\%$ for $N=100$. This performance gain is maintained both for $\psi=0$ and for $\psi=0.25$: by using \gls{libra}, nodes can avoid transmitting low-value updates, and the cost of collisions is more than offset by the higher value of successful transmissions. Additionally, as Fig.~\ref{fig:sens_en} shows, the overall energy consumption (i.e., mean number of transmissions) of \gls{libra} is about $20\%$ lower than that of the \gls{cdns} with $\psi=0$. This advantage decreases in the case with $\psi=0.25$, as the \gls{cdns} solution also avoids transmitting low-value updates, but is still significant when the number of nodes is relatively small. Finally, the energy consumption is evenly distributed among the nodes, while the \gls{cdns} solution places all the load on a single node, depleting its battery at a much faster pace.

We then consider an asymmetric scenario, in which all nodes follow the $\chi^2$ distribution, but the expected \gls{voi} is different for each node, picked randomly and uniformly in the interval $[1-\nu,1+\nu]$. The parameter $\nu\in\mathbb{R}^+$ controls the relative variation among nodes. Fig.~\ref{fig:wsn_asymm} shows the performance of \gls{libra} over $200$ independent realizations. As the expected reward and energy consumption are different for each scenario, we plot the relative gain over the (pull-based) \gls{cdns}. 

First, we consider the relative reward, whose empirical \gls{cdf} is shown in Fig.~\ref{fig:sens_asymm_gain}: \gls{libra} outperforms the best \gls{cdns} in $80\%$ of cases with $\psi=0$ and $90\%$ of cases with $\psi=0.25$. We also note that it is also possible to fall back to the \gls{cdns} when relative performance is lower than $1$, as the expected reward for all \glspl {cdns} can be easily computed in closed form. On average, \gls{libra} outperforms the best \gls{cdns} by slightly more than $5\%$ for $\psi=0$ and $10\%$ for $\psi=0.25$. We observe that the reward gap is smaller than in the symmetric case: as some nodes have a much higher reward (the maximum possible expected reward is $3$ times higher than the minimum), the benefit of knowing the actual value is less significant than the additional coordination cost. However, \gls{libra} maintains another clear advantage, shown in Fig.~\ref{fig:sens_asymm_en}-\subref*{fig:sens_asymm_fair}: the total energy consumption is always between $20\%$ and $30\%$ lower than the \gls{dns} solution, and it is distributed much more fairly among nodes, further improving the lifetime of the \gls{wsn}. In a large majority of cases, \gls{libra} can then deliver a higher expected \gls{voi}, while also significantly reducing energy consumption, even in a favorable scenario for pull-based operation.

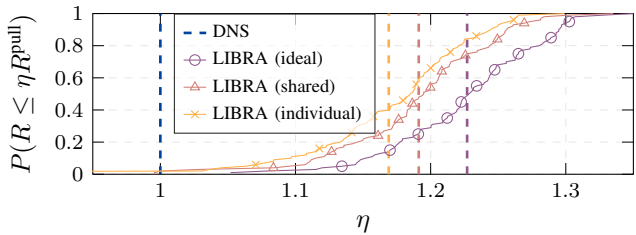
\begin{figure}
\centering
\begin{tikzpicture}
\begin{axis}[
    width=\fwidth,
    height=\fheight,
    xlabel={$\eta$},
    ylabel={$P(R\leq\eta R^{\text{pull}})$},
    xmin=0.95, xmax=1.35,
    ymin=0, ymax=1,
    legend pos=south east,
    ymajorgrids=true,
    xmajorgrids=true,
    grid style=dashed,
    legend style={legend columns=1, font=\scriptsize, anchor=north west, at={(0.15,1)}}
]

\addplot[
    color=color0, line width=1pt, dashed
    ]
    table {
1  0
1  1
    };
\addlegendentry{DNS};

\addplot[
    color=color14, mark=o, mark repeat=10, mark phase=5
    ]
    table[x=gain,y=cdf] {fig/fig_data/robustness_gain_ideal.dat};
\addlegendentry{LIBRA (ideal)};
\addplot[
    color=color14, line width=1pt, dashed, forget plot
    ]
    table {
1.2270  0
1.2270  1
    };

\addplot[
    color=color34, mark=triangle, mark repeat=10, mark phase=5
    ]
    table[x=gain,y=cdf] {fig/fig_data/robustness_gain_sha.dat};
\addlegendentry{LIBRA (shared)};
\addplot[
    color=color34, line width=1pt, dashed, forget plot
    ]
    table {
1.1912  0
1.1912  1
    };
\addplot[
    color=color4, mark=x, mark repeat=10, mark phase=5
    ]
    table[x=gain,y=cdf] {fig/fig_data/robustness_gain_ind.dat};

\addplot[
    color=color4, line width=1pt, dashed, forget plot
    ]
    table {
1.1690  0
1.1690  1
    };
\addlegendentry{LIBRA (individual)};

\end{axis}
\end{tikzpicture}
\caption{\gls{libra} robustness to parameter estimation errors over $100$ episodes in the asymmetric \gls{wsn} scenario with $N=10$ and $\psi=\nu=\eta=0.25$.}\vspace{-0.3cm}
\label{fig:wsn_robust}
\end{figure}

We then examine the performance of the \gls{beta} learning strategy: Fig.~\ref{fig:mab_asymm} shows that the \gls{mab} solution converges to about $99\%$ of the reward obtained by \gls{libra} after approximately $20\,000$ steps. Fig.~\ref{fig:mab_symm}-\subref*{fig:mab_asymm} show the best, worst, and average performance out of $100$ different environments. The plots clearly show that the asymmetric scenario is more difficult than the symmetric one, as there are more edge cases and convergence tends to be slower. This difference is also visible when comparing performance after convergence: while the difference in the average performance is negligible, some asymmetric scenarios represent edge cases in which \gls{beta}'s convergence is much slower, and the performance at convergence does not reach the same level as \gls{libra}. However, the gap between \gls{beta} and \gls{libra} remains relatively small and \gls{beta}  significantly outperforms the best \gls{cdns}.

Finally, we perform a robustness test: we set $\nu=0.25$, and add another error on the knowledge of the nodes, so that the nodes estimate an expected value $\hat{V}_n\sim\mc{U}(\E{V_n}-\eta,\E{V_n}+\eta)$. We consider both a shared case, in which all nodes have the same $\hat{V}_n$ for each node, and an individual case, in which there is an \gls{iid} value of $\hat{V}_{m,n}$ for each node $m$, representing its belief over node $n$. Naturally, \gls{beta} is unaffected by this error, as it does not rely on any prior knowledge.

The performance of \gls{libra} relative to the best \gls{cdns} alternative is shown in Fig.~\ref{fig:wsn_robust}: there is a certain gap between the case with the correct information and the other two, as the average improvement over the pull-based solution goes from $22\%$ to $19\%$ for shared estimates and $17\%$ for individual estimates, but most of the gains are preserved. Furthermore, the inefficiency caused by the imperfect knowledge spreads throughout all individual scenarios instead of becoming catastrophic in some scenarios. Relative rewards below $1$, i.e., scenarios in which \gls{libra} does worse than the pull-based solution, only happened in the individual values scenario, and even then in a very few cases. 

\section{Conclusion}\label{sec:conc}

This paper presents a theoretical model for \acrfull{goma}, an extension of the \gls{goc} paradigm from point-to-point encoding problems to medium access control. To our knowledge, this is the first systematic model of \gls{goma}. We also present \gls{libra}, a low-complexity algorithm that obtains locally optimal solution to the problem, and \gls{beta}, a learning-based solution leveraging the distributed semi-bandit framework to allow nodes to converge to the \gls{libra} solution with limited signaling.

This is a first step toward a complete characterization of \gls{goma}, as it considers a simple collision channel, and assumes nodes have independent \glspl{voi}. 
An important future challenge is  accounting for the time correlation of \gls{voi}: as communication actions affect the estimate of the next measurements, moving from a memoryless model to one that considers the consequences of actions over multiple steps is a key problem for \gls{goc} systems that we will aim to address in future work.

\bibliographystyle{IEEEtran}
\bibliography{IEEEabrv,biblio.bib}

\end{document}